# Recognizing Temporal Linguistic Expression Pattern of Individual with Suicide Risk on Social Media

Aiqi Zhang[1,2], Ang Li[3], Tingshao Zhu[1,4*]

[1]Institute of Psychology, Chinese Academy of Sciences, China

[2]University of Science and Technology of China, Hefei, Anhui, China

[3]Department of Psychology, Beijing Forestry University, Beijing, China

[4]Institute of Computing Technology, Chinese Academy of Sciences, China

*Corresponding author: Tingshao Zhu tszhu@psych.ac.cn

**ABSTRACT**

Suicide is a global public health problem. Early detection of individual suicide risk plays a key role in suicide prevention. In this paper, we propose to look into individual suicide risk through time series analysis of personal linguistic expression on social media (Weibo). We examined temporal patterns of the linguistic expression of individuals on Chinese social media (Weibo). Then, we used such temporal patterns as predictor variables to build classification models for estimating levels of individual suicide risk. Characteristics of time sequence curves to linguistic features including parentheses, auxiliary verbs, personal pronouns and body words are reported to affect performance of suicide most, and the predicting model has a accuracy higher than 0.60, shown by the results. This paper confirms the efficiency of the social media data in detecting individual suicide risk. Results of this study may be insightful for improving the performance of suicide prevention programs.

**INTRODUCTION**

Suicide is one of the leading causes of deaths worldwide. The World Health Organization (WHO) reported that, on average, a suicide occurs in every 40 seconds (World Health Organization, 2014). Furthermore, the effect of completed suicide on families and communities are often devastating (Clark & D Goldney, 2000; Jordan & McIntosh, 2011). Many suicide deaths are preventable (Bailey et al., 2011), which requires to detect individual suicide risk among populations in real-time. However, traditional methods (e.g. self-report ratings, structured interview, and clinical judgment) fail to meet the requirement of early detection (McCarthy, 2010), which suggests the need to improve existing methods.

The development and prevalence of social media sheds new light on this direction. Social media users are motivated to disclose themselves online, and some of them even have broadcasted their suicide thoughts and behaviors on social media sites, such as Facebook and Sina Weibo (Murano, 2014). More importantly, social media data is publicly available and can be processed in real-time. It implies that social media might be an efficient platform for detecting individual suicide risk. Because the characteristics of the language use can be indicative of personal inner process (e.g. thoughts, intentions, and motivations), conducting content analysis on personal linguistic materials (e.g. suicide notes) is recognized as an important method for understanding suicidal thoughts and behaviors (Pestian, et al, 2012; Cheung, et al, 2015). Many studies have confirmed correlations between the frequency of any words in social media posts and individual scores on suicide risk, suggesting the linguistic features of the language use can be used as valid indicators for detecting individual suicide risk on social media (McCarthy, 2010; Sueki, 2015). However, few studies focused on the temporal relationships between the changes in patterns of the language use and levels of suicide risk, which may provide insight into the track of the suicidal process on social media. Li et al (2014) used the Chinese Linguistic Inquiry and Word Count (CLIWC) program to analyze a 13-year-old boy's 193 blogs, which were posted during the year prior to his suicide death. They found that the temporal analysis method can be helpful to understand the suicidal process of completed suicides on social media. Although the temporal analysis on social media data provides an opportunity to identify those at risk of suicide, computational methods of online detection have not yet been fully established (Christensen, et al, 2014). Besides, due to the limited sample size, the performance of the temporal analysis also needs to be tested on a larger population.

This paper aims to examine the temporal patterns of suicidal users' language use and then use of such patterns to build classification models for differentiating people with higher and lower levels of suicide risk.

**METHOD**

Because Weibo (weibo.com) is the most popular Chinese micrologging service provider in China, this study was conducted on such platform. This study consists of two steps: Examining temporal patterns of the suicidal language use and Building classification models for identifying people at risk of suicide.

Examining temporal patterns of the suicidal language use: We examined temporal patterns of the suicidal language use using a three-step procedure: (1) Collecting Weibo posts; (2) Extracting linguistic features; (3) Exporting time series data; and (4) Conducting cluster analysis on time series data.

To identify the unique use of language among suicidal users, we need to collect Weibo posts from individuals with different levels of suicide risk (i.e. individuals who complete suicide, individuals at high risk of suicide, and individuals at low risk of suicide), separately.

To find out users who complete suicide, we contacted with a user (逝者如斯夫 dead), who is famous for collecting news reports about suicidal users. He advised us a list of suicidal users. We confirmed the list by both double-checking relevant news reports and looking through comments left on Weibo accounts of those suggested users. Then, we did a further scrutiny of those confirmed suicidal users to exclude the following: (a) users who are not Chinese citizens; (b) users who might update Weibo posts for business purposes (e.g. movie and sports stars); (c) users who updated less than 20 posts. Finally, we got a total of 30 suicidal Weibo users (15 males and 15 females) and downloaded their Weibo posts since registration.

To find out users at high or low risk of suicide, we randomly selected 1,500 users from a customized Weibo database composed of 1.06 million active Weibo users (Li, et al., 2014). All 1,500 users were administrated by Suicidal Possibility Scale (SPS), which is an effective screening tool designed to assess suicide risk in adolescents and adults (Gençöz, 2006; Naud, 2010). SPS consists of 36 self-rating items. Participants rated themselves on each item by a 4-point Likert Scale (1=None or little of the time to 4=All of the time). High scores indicate high suicide risk. The Chinese version of SPS was used in this study (Huang et al., 2012). The Cronbach's Alphas for the whole-questionnaire was ? in our data. All participants were divided into high risk and low risk group based on scores of the test (Li, et al., 2014). In this study, we randomly selected 30 users from each one group and got a total of 60 users. Then, we downloaded Weibo posts of these 60 users since registration.

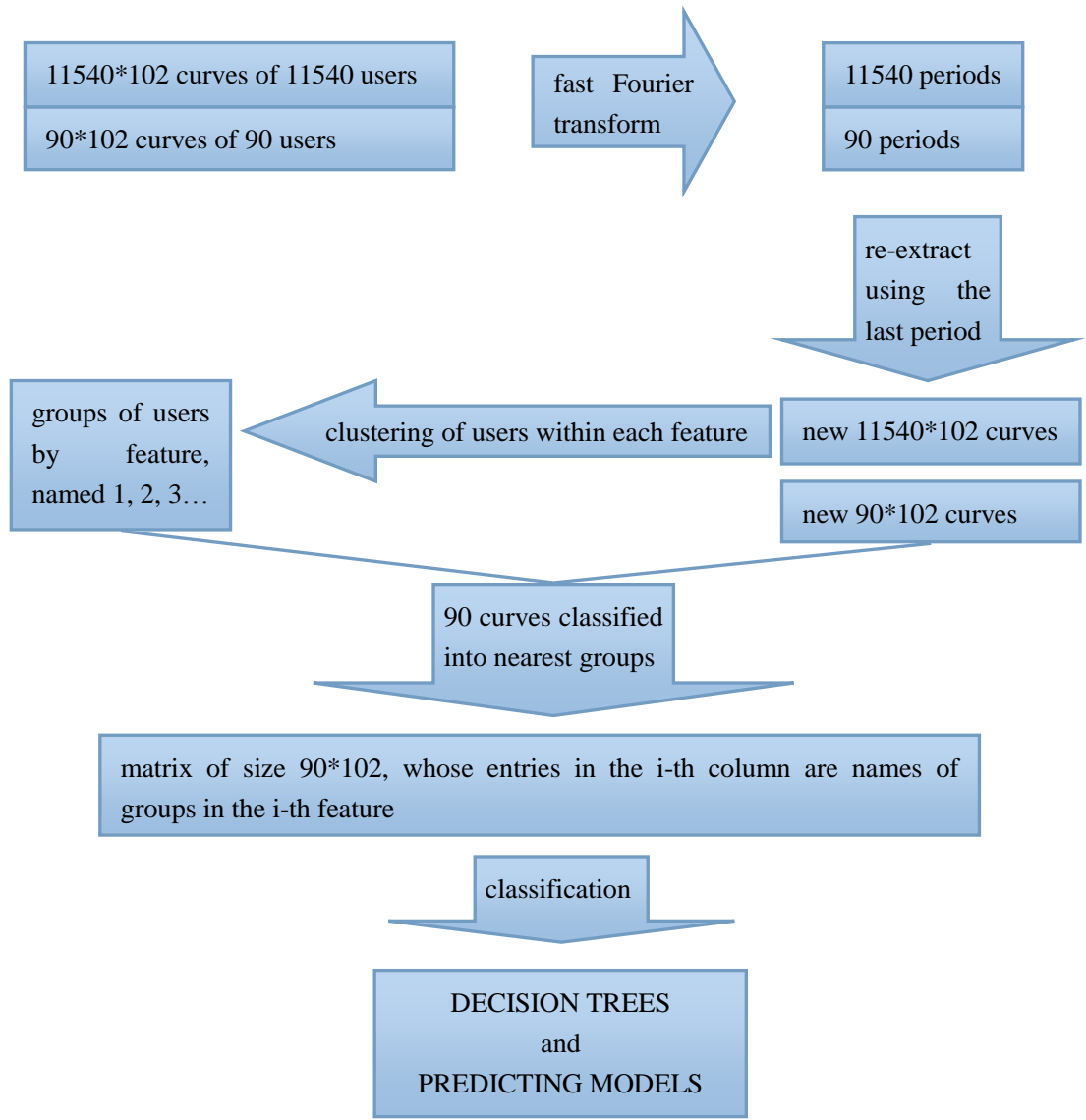

## Pre-processing

We extracted linguistic features from collected Weibo posts using SCMBWC (Simplified Chinese Micro-blog Word Count Dictionary) and TextMind, based on the Chinese version of LIWC (CLIWC). This provides 80 psychologically meaningful language variables (e.g. emotion words) and other 22 linguistic features indicating the frequency of using different punctuation marks and words, such as the number of word-per-sentence and frequency of latin words. Finally, we got a total of 102 linguistic features (80+22). For each user, we computed feature values per day and produced a time vector (102 × number of observation days) composed of 102 time series data. Sequences of 80 linguistic features and 11 kinds of punctuation were obtained by computing frequency of certain features using day as unit, while sequences of the other 11 features described counting of words and word-per-sentence, rates of Latin word, numbers of URLs and so on.

After producing individual time vectors, we found there existed differences in the length of time vectors among different users. For example, among a total of 90 users, the maximum length of the time vector was 1844 (days) and the minimum length was 42 (days). Because some users updated their Weibo posts with few Chinese characters in the last 42 days of the observation period, we cannot make the length of observation uniformly by selecting the minimum length of observation. Besides, because of the different patterns of use, different users have personalized plans for updating Weibo posts. In our dataset, we found that, among different users, the time period for updating the last 100 Weibo posts ranges from 2 days to 2 years. Therefore, we also cannot conduct analysis on a certain number of latest Weibo posts for all users.

To solve the problem, for each time series data, we used the algorithm of Fast Fourier Transform (FFT) to compute the Discrete Fourier Transform (DFT), which was recognized as the amplitudes of all possible frequencies. we viewed changes of values to the features by day as personally fluctuations of mental or physical state and used Fourier transformation to compute the period. A fast Fourier transform (FFT) is implemented to decompose the original time sequence as several sin curves with different frequencies. With this we take periods to the maximal amplitudes for the re-extracting of the time sequences. From each user we intercept the last period of his sequence and use 1/12 of the period as time unit. 1/12 is defined because the shortest period is 12 days. By this method we got sequences with fairly good consistency.

Besides, to cluster and classify the 90 users into clusters, 11,540 users were chosen randomly from a specific group of Weibo users, of whom the total duration since the accounts opened was between 50 - 1800, consistent with that of the 90's. By conducting the same process to all the users, we now obtained the final (11540 + 90)*102 time sequences of length 12 for learning and predicting.

## Smoothing and Fourier transforming

Missing values, referred particularly items equaling 0 in the sequences in our case, meant the user didn't post any words in a certain period of time. The missing data effects statistical inference much, since one didn't post anything online never means that there was no linguistic behavior.

To deal with zero values in the sequences, we applied lpint (Martingale estimating equation local polynomial estimator of counting process intensity function) in package lpint to the sequences. Take social-life words (SLW), one of the typical patterns mentioned above, as an example, we model the numbers of SLW by a doubly stochastic Poisson process N(t) with intensity process Y(t)a(t), where Y(t) counts the words of Weibo posts at time t, and a(t) denotes the SLW rate at time t. We estimated the intensity function with the local polynomial intensity estimator proposed in (Chen, Yip, and Lam 2011), and the estimator for a(t) is given by

$$\hat{\alpha}(t) = e_1^T A(t)^{-1} \int_0^1 K_b(s-t)g(s-t)\frac{J(s)}{Y(s)}dN(s),$$

where

$$A(t) = \int_0^1 K_b(s-t)g(s-t)g(s-t)^T ds$$

$$J(s) = I_{\{Y(s)>0\}},$$

with the convention J(s)/Y(s) = 0 when Y(s) = 0 to protect against division by zero, and with the notations introduced as follows. $e_i = (0, ..., 0, 1, 0, ..., 0)^T$ is a (p+1)-dimensional unit vector with 1 as its i-th component, $g(x) = (1, x, ..., x^p/p!)^T$ and $K_b(\cdot) = (1/b)K(\cdot/b)$ is the scaled version of the Epanechnikov kernel where bandwidth b is decided by a data-driven selector.

We got an output of a smoothed vector with length longer than length of the input vector by 2. The smoothed sequence was obtained by excising the sub-sequence of the second to the last but one items. By normalizing it we got the final sequence for later processing, while the same method was implemented (11540 + 90) times, covering every user studied. (11540 + 90) smoothed and normalized sequences were obtained.

Then we applied Fast Fourier Transformation again to the smoothed sequence to get the period properties of them. For each of the 102 features, we combined descriptive statistics (mean values and standard deviations) with fft coefficients and phase corresponding to the maximum amplitude, which are extracted from the returned values of the fft function, and got a matrix telling characteristics about the given feature.Then we had a number of rows, each represented one of the users with information of time sequence. So in all there were 204 matrices computed, 102 of them based on the 11540 random users and other 102 based on 90 observations with labels 0,1 and 2.

## K-means clustering

The 102 matrices of row number 11540 were for k-means clustering in our case. For each feature, a proper value for the number of clusters k was found out by respectively calculating Silhouette coefficient while k was possibly taken value between 1 and 7. We then applied function kmeans to matrices and got 102 vectors with levels 1 to the selected number of k, each of length 11540, describing nominally which cluster the subjects were in. Correspondingly we got 102 matrices of cluster centers also.

Matrices telling properties of sequences of observations with labels were subtracted from the centers obtained from kmeans clustering. Absolute values are calculated to show distances. We classified each user into a certain cluster, which was noted by a number. From this we got a 90*102 matrix with every column nominal scales indicating several classes of time sequences. By binding factor vector with levels 0,1 and 2 and length 90 with the 90*102 matrix we obtained a matrix with 103 columns, the last column dependent variable. On this basis, classification could be implemented and decision trees were to be drawn.

## Drawing decision trees and predicting

We aimed at getting the relevant features to the labels and predicting by classifying and computing decision trees. Two method were applied to do this work: rpart (Recursive Partitioning and Regression Trees) in package rpart and train (Fit Predictive Models over Different Tuning Parameters) gbm (Generalized Boosted Regression Modeling) in package caret.

In the rpart method, we draw regression tree based on random 3/4 of all the 90 observations with the not chosen 1/4 used as test set, while by functions in package caret we remove variables whose variance too small and apply function rfe to the training set for feature selection before modeling, sampling as well. 3/4 random users for training and the rest 1/4 for testing.

Different from the first method, we set resampling method, the number of folds and the number of complete sets of folds for cross validation and set the range of parameters needed in random forest algorithm when building boosted models. The output of the modeling were selected variables, confusion matrix out from the predicting process compared with the test set, resampling results and computed relative influence of each variable in the gbm object, generated by function summary.gbm.

From the two classification model we predicted the level of any unstudied time sequence of Weibo users, corresponding to the factor vector label with levels 0, 1 and 2. Variables with the most significant influence were also obtained, by running several times and finding the most frequently appearing features in the summary part.

**RESULTS**

Since training set was random extracted, we got different models with different accuracy every time. Here is a confusion matrix of the learning process using 10-folds.

```
          Reference
Prediction   0   1   2
         0   2   0   1
         1   0   3   1
         2   1   0   1
```

Accuracy : 0.6667
95% CI : (0.2993, 0.9251)
No Information Rate : 0.3333
P-Value [Acc > NIR] : 0.04242
Kappa : 0.5

Figure 1-5 are decision trees generated by **rpart** method, followed by confusion matrices interpreting the effect of predicting.

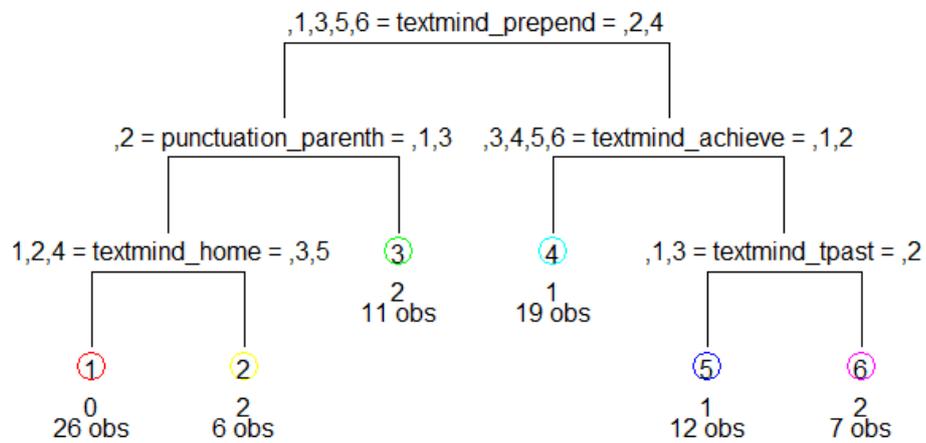

---

Confusion Matrix and Statistics

```
          Reference
Prediction  0  1  2
        0   2  1  0
        1   0  3  0
        2   1  0  2
```

Accuracy : 0.7778
95% CI : (0.3999, 0.9719)
No Information Rate : 0.4444
P-Value [Acc > NIR] : 0.04635
Kappa : 0.6667

---

To study which of the features influences most to the result, the following shows 3 independent experiment of **gbm** fitting and **summary.gbm**.

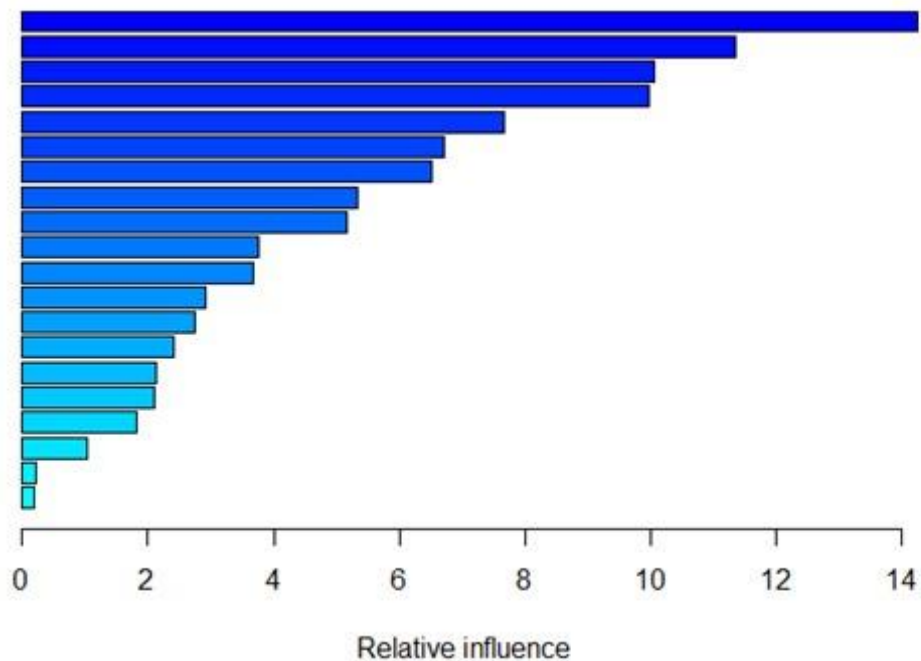

---------------------------------------------

First 5 variables in the summary graph:
body words, auxiliary verbs, motion words, parenthesis, and time words

---

By counting frequency of variables during the computed relative influence, we get 10 most significant ones:

**punctuation_parenth, textmind_achieve, textmind_anx, textmind_auxverb, textmind_body, textmind_feel, textmind_home, textmind_insight, textmind_motion, textmind_ppron**

### Discussion & Comparison with previous case studies
Previous studies worked on the correlation between suicide and LIWC features, selecting features with small p values as significant ones (M. Fernández Cabana, 2012). Other reports show comparison of factors like time career age and word type and conclude the most related treatment (Stirman, 2001). Logistic regression is applied to some data as the independent variables were viewed continuous (Masuda1, 2013). There are no studies on suicide dealing with a bunch of users making use of LIWC before. In our case we classify the observations into several groups by features and draw decision trees using group names made in classifying. Since all the features expressed as nominal scales, there was no correlation test to be computed, and instead of p-value showing the relationship between suicide and variables we obtained relatively most significant ones in among all the features and statistics. Future studies, therefore, can start up in logistic regression with Fourier coefficients of the sequences as independent variables.

Chinese materials were studied in our case, whereas most previous studies have only analyzed materials written in or translated into English, while the writer's original psychological meaning and linguistic style may not be completely preserved after translation (Tim M.H. Li, 2014). Our study differs from previous ones using Chinese version of the LIWC in that we studied a number of social network site users and computed decision trees. Chinese-specific categories may further reveal psychological cues of suicide that cannot be used in other languages (Tim M.H. Li, 2014).

Notice that we use cluster centers to notify properties of clusters, thus these centers should be showed to demonstrate distinctive features for suicide groups. According to decision trees gained from different training sets, among the 10 most significant features, we got 4 relative to group 0, that is, users who committed suicide.

**punctuation_parenth**

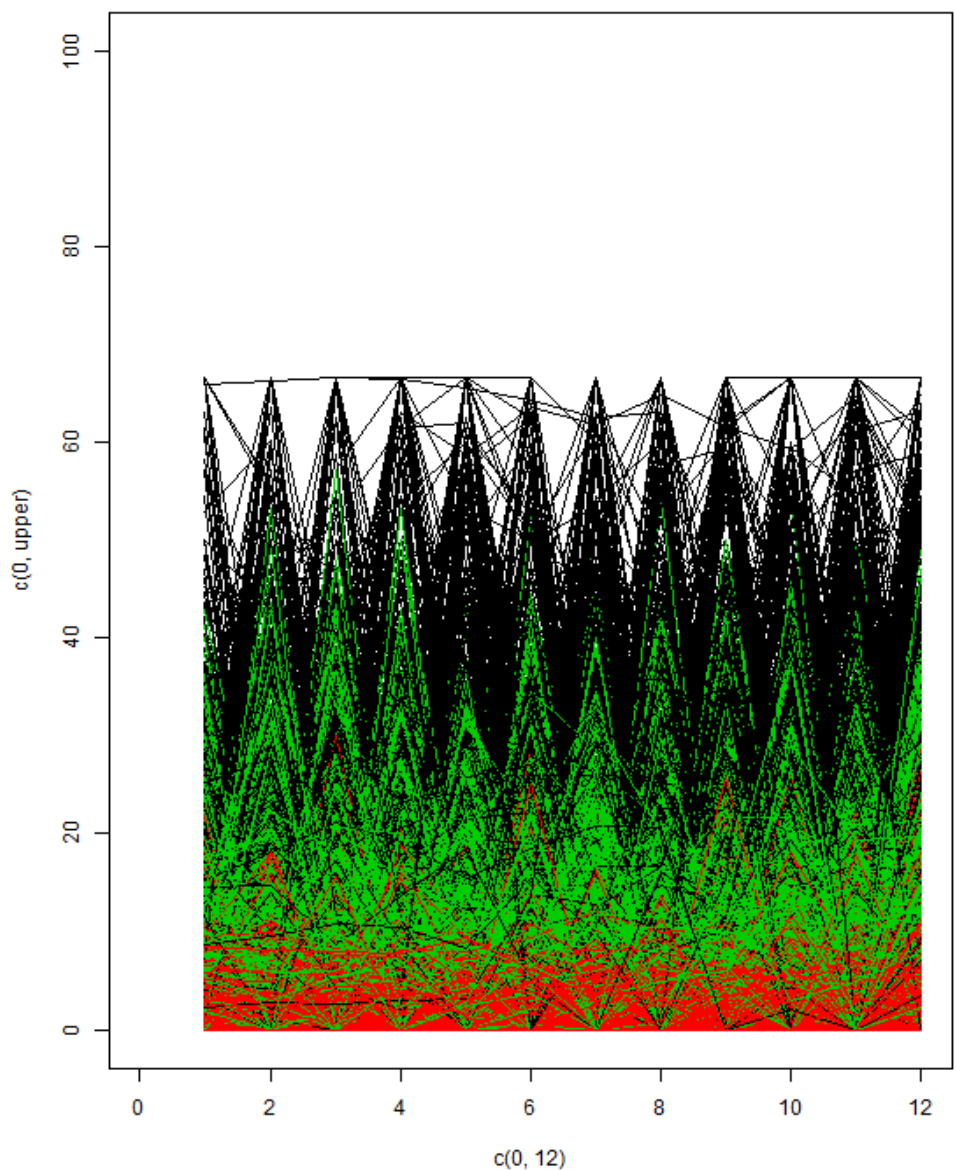

| AVG | SD | 0 | 1 | 2 | Arg1 |
|---|---|---|---|---|---|
| 115.13012161 | 14.51798345 | 0.395844657 | 0.287937124 | 0.205698385 | 0.031531324 |
| 22.741153902 | 2.866167462 | 0.382331809 | 0.310391627 | 0.192305825 | 0.019072178 |
| 38.239399639 | 6.85765465 | 0.42652605 | 0.295963262 | 0.197711739 | 0.033835113 |

**textmind_auxverb**

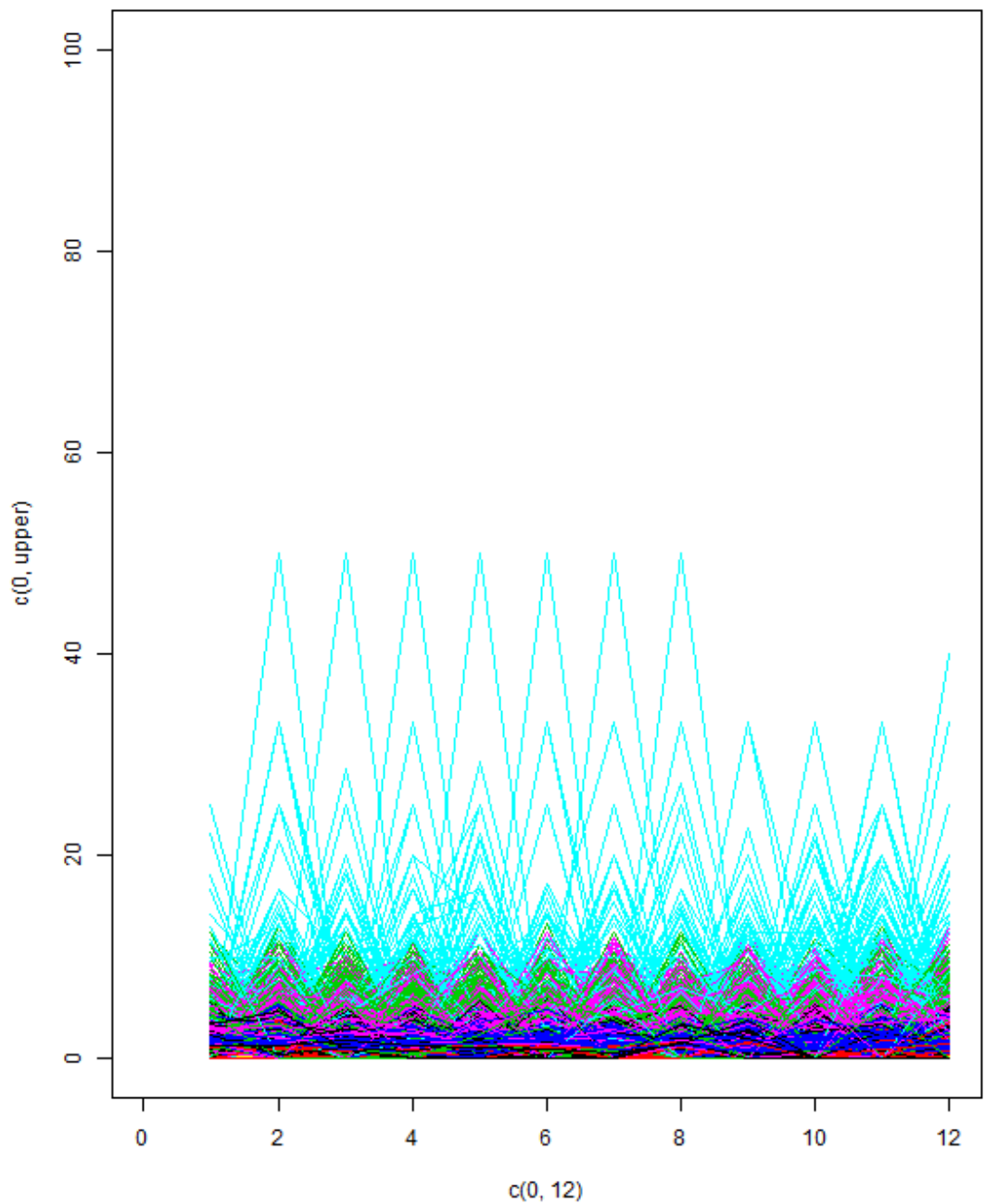

|   | AVG | SD | 0 | 1 | 2 | Arg1 |
|---|---|---|---|---|---|---|
| 1 | 0.859780602 | 1.13359216 | 0.42544777 | 0.3513784334 | 0.192942227 | -0.030703143 |
| 2 | 0.45646321 | 0.643238929 | 0.3817371230 | 0.3604214570 | 0.198317093 | -0.028562451 |
| 3 | 1.23507966 | 2.083186149 | 0.3646119510 | 0.3305475310 | 0.219752759 | -0.016703396 |
| 4 | 1.5690810520 | 0.761592293 | 0.4953067020 | 0.2736044390 | 0.168876502 | 0.101621611 |
| 5 | 2.7462569235 | 5.222238298 | 0.3008070370 | 0.3074088810 | 0.230586356 | 0.007398932 |
| 6 | 2.288426194 | 2.679129902 | 0.4575846830 | 0.3056317840 | 0.191989699 | 0.044695307 |
| 7 | 0.035452023 | 0.099684427 | 0.09538323 | 0.1277912170 | 0.076010697 | 0.011927493 |

**textmind_ppron**

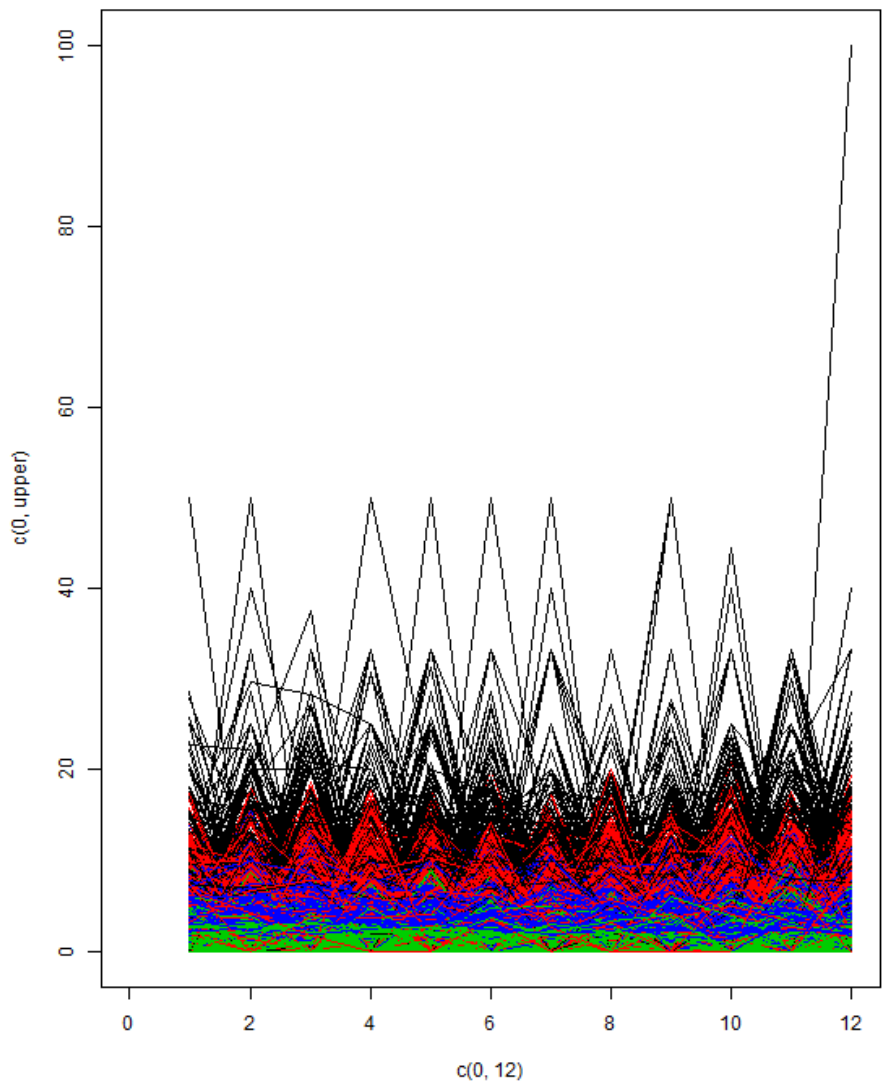

|   | AVG | SD | 0 | 1 | 2 | Arg1 |
|---|---|---|---|---|---|---|
| 1 | 4.886485982 | 5.88412429 | 0.390026292 | 0.309999442 | 0.20601711 | -0.006538162 |
| 2 | 2.605513666 | 3.248422761 | 0.4117724 | 0.320524585 | 0.206776553 | -0.002718911 |
| 3 | 1.357334981 | 1.392897273 | 0.410696572 | 0.324519575 | 0.187144587 | 0.005262552 |
| 4 | 3.547252654 | 1.608387298 | 0.497152643 | 0.275347883 | 0.168847521 | 0.063698304 |

**textmind_body**

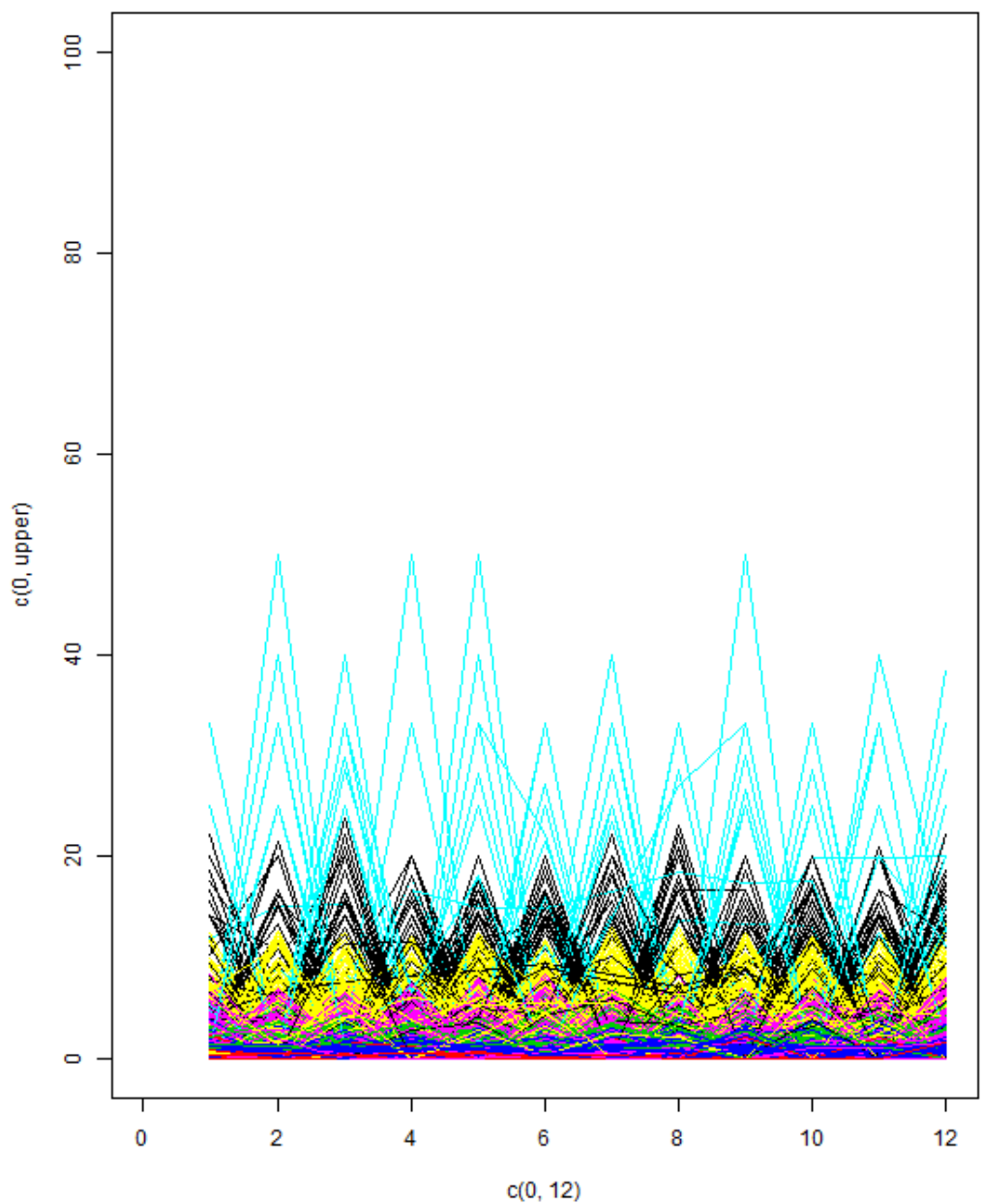

|   | AVG | SD | 0 | 1 | Arg1 |
|---|---|---|---|---|---|
| 1 | 2.395603793 | 4.59783583 | 0.301782598 | 0.318970376 | 0.037743009 |
| 2 | 0.129525066 | 0.263360931 | 0.228623388 | 0.266353604 | −0.012259764 |
| 3 | 1.492920946 | 0.917740263 | 0.488347043 | 0.288160849 | 0.105404415 |
| 4 | 0.649327344 | 0.671164326 | 0.450879339 | 0.338566756 | 0.009294324 |
| 5 | 3.877760453 | 9.115389734 | 0.219022062 | 0.296123209 | 0.038607001 |
| 6 | 0.894588789 | 1.546070027 | 0.362208628 | 0.344195046 | −0.012441021 |
| 7 | 1.666291015 | 2.568380012 | 0.360546753 | 0.324560843 | 0.027588917 |

From the decision trees we see that in the use of parentheses, the suicide group are matched with time sequences in cluster 2. Relatively low average value, small standard deviation and phases corresponding to max amplitudes near 0 are mainly observed to describe the characteristic of this cluster of curves, among which phases near 0 mean that the sine waves with the largest amplitudes start at somewhere near time 0. For use of auxiliary verbs, group of suicides were found high at both FFT coefficient of 1 Hertz and average value (see at parameters of cluster 2, 4, 5, 6 and 7). The large coefficients corresponding to 1 Hertz may imply that the time sequence curves are of relatively significant property of forming one whole sine period.

Besides parentheses and auxiliary verbs we got personal pronouns also relative to classifying groups. All the 4 clusters except the third one are divided to node with suicide group. Comparing parameters we found that cluster 3 has got the lowest level of average value and the smallest standard deviation, which means the suicide group is of relatively much use of personal pronouns, which is in line with previous reports. Another result different from former studies is the use of body words. Cluster 3, 4, 5 and 7 are claimed to be related to suicide group and most curves in these clusters share a similarly high level of phase to the max amplitude. Also coefficients of 0 Hertz are distinctively larger than the other two matched groups, showing that tiny fluctuations are recorded in the curves of suicide group.

**CONCLUSION**

Relationship between suicide and linguistic performance has been studied for long. Previous cases used several methods, including method to treat posting frequency, and the studies focused on materials in several languages, on and off-line. We didn't get the consistent results with former cases except use of pronouns. However there are limitations in our study. 90 is a rather small sample size for classifying users with known labels 0, 1 or 2, and high-score group and low-score group might not be exactly classified. Besides, the CLIWC categories shows only the originally used meanings of the words with no other meanings like irony or sarcasm, not to mention the meaning of the contents.

Still, it is an interesting field for future studies to deal with linguistic and behavioral characteristics of suicides. Our work provides a new method to see into the problem and gives different results from previous studies, which may help to the prediction and cure of psychological health.

**REFERENCE**

Bailey, R.K., Patel, T.C., Avenido, J., Patel, M., Jaleel, M., Barker, N.C., Khan, J.A., Ali, S., Jabeen, S., (2011). Suicide: current trends. J. Natl. Med. Assoc. 103, 614–617.

Cheung, G., Merry, S., & Sundram, F. (2015). Late-life suicide: Insight on motives and contributors derived from suicide notes. Journal of affective disorders, 185, 17-23

Christensen, H., Batterham, P., O'Dea, B., 2014. E-health interventions for suicide prevention. Int. J. Environ. Res. Public Health 11, 8193–8212.

Clark, S. E., & Goldney, R. D. (2000). The impact of suicide on relatives and friends. The international handbook of suicide and attempted suicide, 467-484.

Fernández-Cabana, M., et al. (2013). "Suicidal Traits in Marilyn Monroe's Fragments." Crisis 34(2).

Gao, Rui, et al. "Developing simplified Chinese psychological linguistic analysis dictionary for microblog." Brain and Health Informatics. Springer International Publishing, 2013. 359-368.

Gençöz TOP. (2006). Associated factors of suicide among university students: importance of family environment. Contemporary Family Therapy 28: 261-268. DOI: 10.1007/s10591-006-9003-1.

Guan, L., et al. (2014). Identifying Chinese Microblog Users with High Suicide Probability using Internet-based Profile and Linguistic Features: Classification Model.

Huang C, Chung C, Hui N, Lin Y, Seih Y, Lam B, Pennebaker J. (2012). The development of the Chinese Linguistic Inquiry and Word Count Dictionary. Chinese Journal of Psychology 55: 185-201.

Jordan, J. R., & McIntosh, J. L. (2011). Suicide bereavement: Why study survivors of suicide loss. Grief after suicide: Understanding the consequences and caring for the survivors, 3-17.

Li L, Li A, Hao B, Guan Z, Zhu T. (2014). Predicting active users' personality based on micro-blogging behaviors. PLoS ONE9: e84997 DOI 10.1371/journal.pone.0084997.

Li, T. M., et al. (2014). "Temporal and computerized psycholinguistic analysis of the blog of a

Chinese adolescent suicide." Crisis 35(3).

Masuda, N., et al. (2013). "Suicide ideation of individuals in online social networks." 9(1).

McCarthy, M. J. (2010). Internet monitoring of suicide risk in the population. Journal of affective disorders, 122(3), 277-279.

Murano G. (2014). 8 shocking suicide attempts posted on the Internet. Available at http://www.oddee.com/item 98907.aspx (accessed 8 July 2015).

Naud H, Daigle M S. (2010). Predictive validity of the Suicide Probability Scale in a male inmate population. Journal of Psychopathology and Behavioral Assessment 32: 333-342. DOI: 10.1007/s10862-009-9159-8.

Pestian, J. P., Matykiewicz, P., Linn-Gust, M., South, B., Uzuner, O., Wiebe, J., ... & Brew, C. (2012). Sentiment analysis of suicide notes: A shared task. Biomedical informatics insights, 5(Suppl 1), 3.

Stirman, et al. (2001). "Word use in the poetry of suicidal and nonsuicidal poets." Psychosomatic Medicine 63(4).

Sueki, H. (2015). The association of suicide-related Twitter use with suicidal behaviour: A cross-sectional study of young internet users in Japan. Journal of affective disorders, 170, 155-160.doi: 10.1016/j.jad.2014.08.04

Tausczik, et al. (2010). "The Psychological Meaning of Words: LIWC and Computerized Text Analysis Methods." Journal of language and social psychology 29(1).

World Health Organization, (2014). Preventing Suicide: A Global Imperative. World Health Organisation, Geneva, Switzerland.